# Social media and mobility landscape: uncovering spatial patterns of urban human mobility with multi source data

## Social media and mobility landscape

Yilan Cui[a], Xing Xie[b], Yi Liu[a]*

[a] School of Environment, Tsinghua University, Beijing, 100084, China

[b] Microsoft Research Asia, Microsoft Corporation, Beijing, 100080, China

*Corresponding author: yi.liu@tsinghua.edu.cn

## Graphic abstract

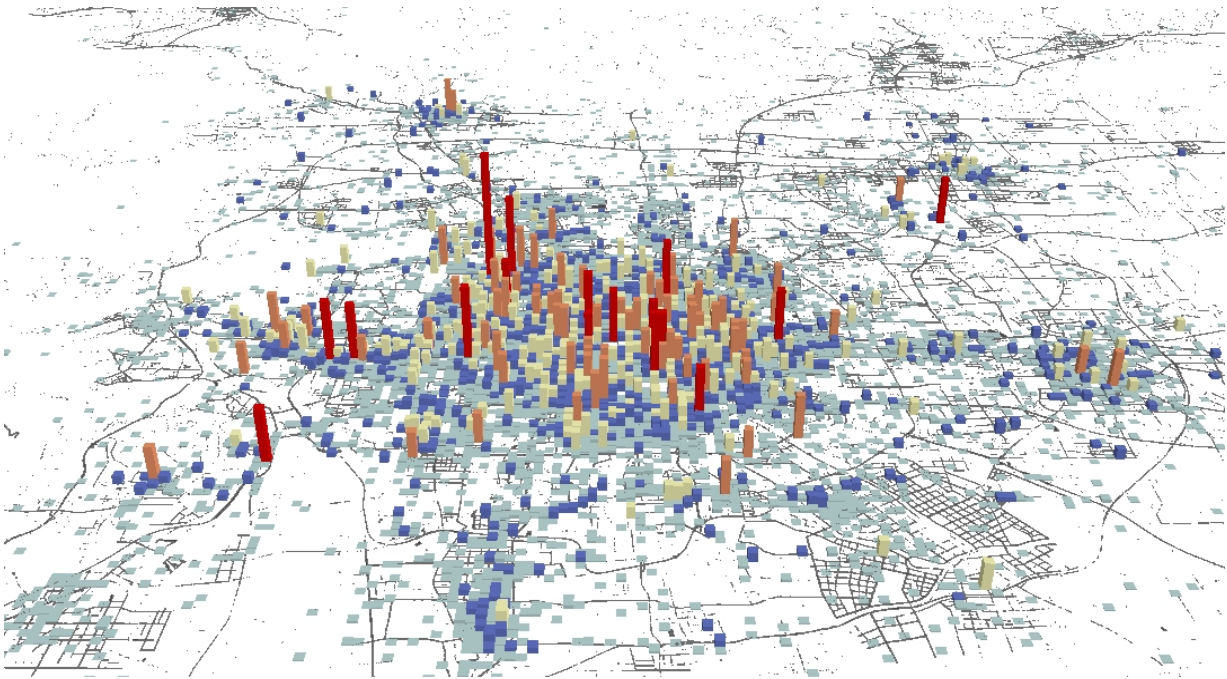

## Highlights:

·   Identify personal activity-specific places based on Weibo data and surveys

·   Propose ways for detecting and moderating sample bias of Weibo data

·   Present a graphic representation of urban activity intensity in Beijing, China

·   Introduce the potential application of Weibo data for urban analysis

**Abstract:** In this paper, we present a three-step methodological framework, including location identification, bias modification, and out-of-sample validation, so as to promote human mobility analysis with social media data. More specifically, we propose ways of identifying personal activity-specific places and commuting patterns in Beijing, China, based on Weibo (China's Twitter) check-in records, as well as modifying sample bias of check-in data with population synthesis technique. An independent citywide travel logistic survey is used as the benchmark for validating the results. Obvious differences are discerned from Weibo users' and survey respondents' activity-mobility patterns, while there is a large variation of population representativeness between data from the two sources. After bias modification, the similarity coefficient between commuting distance distributions of Weibo data and survey observations increases substantially from 23% to 63%. Synthetic data proves to be a satisfactory cost-effective alternative source of mobility information. The proposed framework can inform many applications related to human mobility, ranging from transportation, through urban planning to transport emission modelling.

## 1 Introduction

Urban passenger transport is leading to a range of societal and environmental externalities including congestion, resource consumption, greenhouse gas emissions, and localized pollution. Travel demand is derived from individuals' need for performing activities at different geographic locations. A current trend of research in the field of human mobility is activity-based analysis (Zhang et al., 2013; Rieser et al., 2014; Wu et al., 2014), which takes into account individual's diverse travel demands behind spatial movements. Broadly stated, activity-based approaches pay close attention to the responses to the common questions that each individual faces on a daily basis: 'what activity will I perform?', 'with whom?', 'when?', 'for how long?', 'how do I get there?' and so on (Ghauche, 2010). Such approach allows researchers to examine, in detail, the diversity of travel demands that spur movements, thus to dig into the driving forces of transport energy consumption and emissions. Therefore, identifying the purposes behind human movements becomes a crucial part of understanding human mobility. Such analysis relies heavily on the quality and representativeness of human space-time data, which is traditionally obtained through travel surveys. However, data collection is extremely time-consuming and expensive, which in turn limits its population coverage. Furthermore, such surveys are not carried out on a continuously temporal basis. For example, the most authoritative citywide travel survey in Beijing, China is only conducted every five years.

Over the past decade, the introduction and penetration of location-based services have profoundly changed the way people live. People leave digital traces of their movements everywhere, when swiping smart cards on public buses, making phone calls with friends, or sharing activity related choices on online social networks. This large-scale user-generated data provide new perspectives for spatiotemporal analysis of travel behavior. Temporal, spatial, and textual information in this data are revealed by users in realistic situations, which makes them less prone to cheating. Since such data have already been collected to help operate systems or provide user-side services, additional uses for analysis incur little marginal cost. Different from Call Detail Records (CDR) (Isaacman et al., 2011), Smart Card Data (SCD) (Yuan et al., 2013; Zhou and Long, 2014), and trajectories derived from GPS trackers (Zhang et al., 2013), check-in data have two unique features, which make them more suitable for activity-based mobility analysis. First, check-in records not only contain geo- and time-tagged locations but also include the Point of Interests (POI) categories of the venues indicating the potential functionalities of such locations (Hasan et al., 2013; Wu et al., 2014). Second, users' sociodemographic profiles are available, enabling comparative analysis of travel behavior for different social groups. Recently, a number of researchers have been using check-in data to understand and model how individuals move in time and space (Hasan et al., 2013; Wu et al., 2014; Hasan and Ukkusuri, 2014; Yang et al., 2015; Wang et al., 2015; Chen et al., 2016). However, these studies limit the understanding of mobility dynamics due to the lack of attention to the purposes behind human movements. Hasan et al. (2013), Wu et al. (2014), and Hasan and Ukkusuri (2014) simply identified activity based on the POI category of each geo-tagged location. Potential problems arise from this identification method. For example, checking in at a shopping mall does not necessarily indicate engaging in recreational activities; it may also be a shop manager complaining about his/her tedious day. Besides, each time users post a microblog, they have the option to choose a location to share from several nearby locations. Out of privacy and safety concerns, people tend to provide misleading information.

Previous studies have demonstrated that human mobility shows a high degree of regularity (Hasan and Ukkusuri, 2014; Wu et al., 2014; Wang et al., 2015). Far from randomness, people spend much of their time at a few important places, such as home and workplace, where they visit regularly and often during the same period of the day (Isaacman et al., 2011). An individual's movements typically will either be centered around home, work, or somewhere in between the two locations as they commute in between them (Cho et al., 2011). Some studies have proposed suitable algorithms for predicting semantically meaningful places in somebody's life (especially home locations), based on the spatial and temporal features of movement records. The most commonly used method is a 'Most Check-ins' method. Home is assumed to be the location from which people most frequently tweet (Pontes et al., 2012; Bojic et al., 2015; Hossain et al., 2016), (or make the maximum number of phone calls in the case of CDR (Isaacman et al., 2011), or most frequently visit in the case of SCD (Yuan et al., 2013)). In some cases, the sequence of a location in a user's daily movements (Zhou and Long, 2014; Hossain et al., 2016) or surrounding land use features (Zhou and Long, 2014) is used to determine the likelihood of corresponding activity.

Despite the fascinating features mentioned above, it has been argued that the limited population representativeness would confine the scope of big data research on human mobility (Wu et al., 2014; Chen et al., 2016). Travel survey data, acquired through rigorous sampling methods, is considered as a relatively representative sample of the real population. However, only limited studies have conducted out-of-sample validation of big data with such a dataset, quantitatively analyzing the effect of representativeness of big data (Lenormand et al., 2014; Toole et al., 2015; Yang et al., 2015). Most of the time, validation is performed on a coarse-grained spatial scale, such as municipality (Lenormand et al., 2014) or town (Toole et al., 2015), because raw census data with high spatial resolution remains inaccessible to the public. It remains an interesting question to ponder whether social media check-in data would be suitable for regional activity-based travel analysis, whether and what mobility measures are sensitive towards sample bias (Chen et al., 2016), and most importantly, how to modify such bias.

Population synthesis is a widely-adopted technique for creating a full population microdata based on limited number of observations, to enable analyzing estimates of variables at different spatial scales. It was first developed in the field of economics in the 1960s, and has since been introduced to the fields of geography and social sciences (Hermes and Poulsen, 2012; Ma et al., 2014), for example, to analyze social policy and population changes (Haase et al., 2010), transportation (Beckman et al., 1996; Zhang et al., 2013; Ma et al., 2014), building energy consumption (Subbiah et al., 2013) and so on. In the transport-related field, the pioneering attempt of population synthesis goes back to 1996, when Beckman et al. (1996) applied Iterative Proportional Fitting (IPF) to create a baseline synthetic population of individuals and households, so as to estimate future travel demand. After years of implementation and testing, population synthesis has become a popular and cost-effective way for creating disaggregated data for spatial analysis and simulation, generating long-term forecasts, and examining the geographical effects of government policies (Zhang et al., 2013; Subbiah et al., 2013; Ma et al., 2014). In this article, we explore the potential application of this technique for sample bias modification. A critical review on current algorithms to generate synthetic spatial microdata can be found in Hermes and Poulsen (2012) and Müller and Axhausen (2010).

In this study, we introduce how to identify home, work/school, and destination choices for non-commuting activities based on multi source data in a simple and straightforward way and apply classical methods of population synthesis to solve the bias-modifying problem of social media data in a provably promising way. Compared with existing studies on activity-based mobility analysis with check-in data, we go one-step further to detect and moderate the impact of sample representativeness on mobility measures at sufficiently high spatial granularity.

## 2 Materials

### 2.1 Study area

As the capital city of China, Beijing was selected as our case study area. The city can be divided into four zones including inner area, functional extended area, new urban district, and ecological conservation area (Fig. 1). The inner-city districts of Dongcheng and Xicheng represent the traditional business districts. The functional extended area includes the Chaoyang district in the northeast (which is home to the Beijing International Airport, and Beijing's growing central business district), Haidian in the northwest (which was deliberately developed as a university area, with most of the universities, research institutes, and hi-tech firms located there), Fengtai in the southwest (which is the interflow center of goods in the southwest Beijing), and Shijingshan in the far west (which is the heavy industry center in this city) (Ma et al., 2014).The new urban district composes Tongzhou in the east (where a new administrative sub-center is emerging), as well as districts of Shunyi, Daxing, Changping, and Fangshan. The ecological conservation area refers to the remote counties and villages.

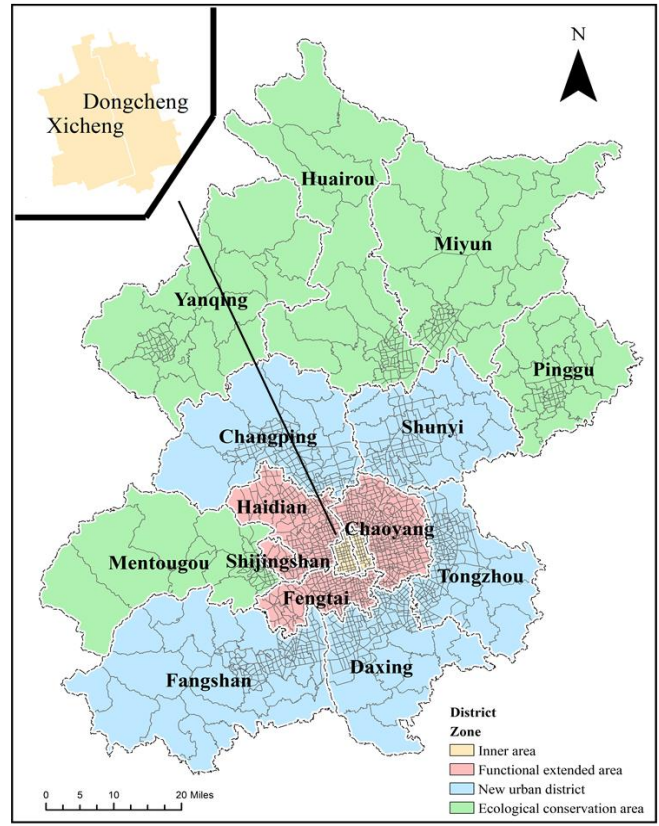

Fig. 1 Municipal districts of Beijing. The smallest geographical units in this figure represent the Traffic Analysis Zones.

Traffic analysis zone (TAZ) is the unit of geography most commonly used in travel surveys. In the Forth Survey of Citywide Travel Logistics of Beijing, the metropolitan area was divided into 1,911 TAZs, with the zonal area ranging from 0.13km$^2$ to 382.03km$^2$, among which 87% is smaller than 5 km$^2$.

**2.2 The mobility datasets**

**(1) User Profiles and Location Check-ins:** This dataset contains 11,961,502 geo- and time-tagged check-in records from 135,736 venues in Beijing by 1,670,968 registered Sina Weibo users from Mar. 2011 to Sep. 2013, which was crawled through Sina Weibo API (Zhang et al., 2012). Each check-in contains the user ID, check-in time, the venue's geo-coordinates, and POI category. When registering for an account in Weibo, users can choose to share some of their personal information on their home pages, including gender, birthday, education background, and marital status *etc*. After removing users with less than 15 check-ins, we obtain 7,324,014 check-ins at 111,955 venues from 161,015 users where each user has 45 check-in records on average.

**(2) Fourth Survey of Citywide Travel Logistics of Beijing:** This survey tracks more than one hundred thousand respondents' socio-economic attributes, as well as their detailed 24h travel diaries in a generic workday or weekend, including a list of activities participated and corresponding trips along with temporal, spatial and travel mode information.

**2.3 User profiles**

In our Weibo sample, almost all the users provide their gender information, 46% users provide age information, and 53% users provide educational background. We found that Weibo is more popular among females, young people and people with high levels of education. Obvious differences exist in the age and education profiles of population covered by these two datasets (*c.f.* Fig. 2); whereas, only a slight difference of 4% is discovered in the gender profiles.

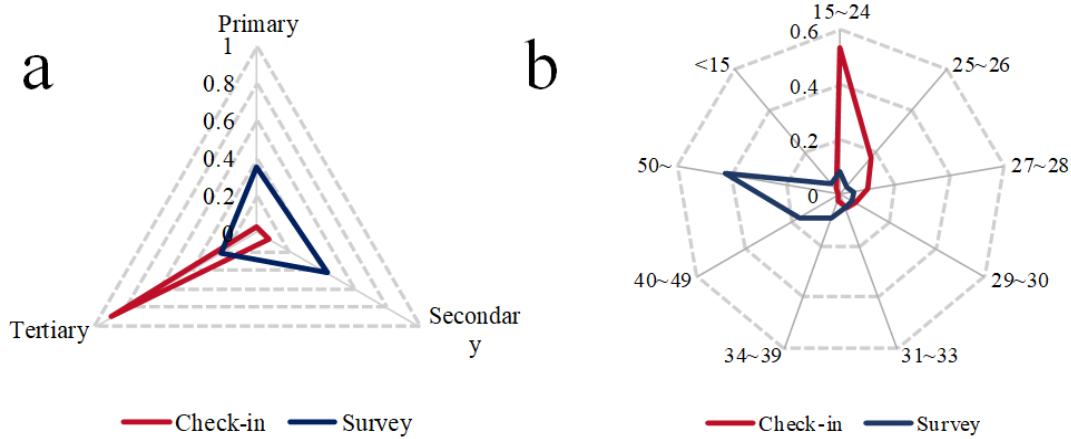

Fig. 2 User profiles. (a) Education background, (b) Age. Here, education background is classified into three categories; including primary (junior school and below), secondary (senior high school), and tertiary (college and above).

**2.4 Categorizing travel demands**

To ease the analysis, daily activities are first classified into a limited number of categories. In this research, considering the temporal characteristics of travel demand inferred from trip records in survey data (*c.f.* Appendix A), we group activity-travel purpose into home, work (work/school), entertainment (eating out, shopping and other leisure activities), and other (personal business, business trip, visiting friends/relatives, *etc.*).

**3 Methods**

**3.1 Ground truth for home and workplace identification**

Building the ground truth is challenging, because it involves identifying a Weibo user's home and workplace from several locations the user checked-in without being told. Intuitively, if someone visits the same residential building frequently, this place holds a big change to be a living place. Similarly, an office building where someone recursively checks-in, is more likely to be a working place. Besides, reading a Weibo content that says 'enjoying my new working station', one can easily tell that it might be sent from a workplace. Weibo content that says 'Good night, Beijing' is most likely sent from home. In this research, we rely on POI category and content information together to build the ground truth for the home and work location identification algorithm.

Firstly, we construct the ground truth for home by first filtering the locations each individual checks in for 4 days or above long, and with a POI category of residential place. Similarly, ground truth for workplace is filtered from locations each individual checks in for 4 days or above long, and with a POI category of corporation, school or industrial park. Then location with largest check-in days in each case is labeled as 'home' and 'work' respectively.

Secondly, to improve the reliability, we further crawl the Weibo contents posted from these labelled locations, linked by user ID and geo-coordinates. Then, we select a set of 20 keywords (*e.g.* 'home', 'dormitory', 'sleep', 'office', 'work' *etc.*) which are most likely to be mentioned in Weibo contents sent from home or workplace (*c.f. SI* Table 1). Next, based on these keywords, we label 'home' or 'work' tag to each user's potential locations. For example, if at least one of the home-related words is contained in any of the Weibo contents posted from user's filtered locations, then this location would be labeled as 'home'.

Finally, we only retain the filtered locations, which have been tagged the same label from the two steps.

As a result, we obtain a training dataset containing 90,284 check-ins from 377 users, among which 296 users' homes and 180 users' workplaces are labelled. Here, the labeled location is represented by the geo-coordinates of the corresponding venue. As can be seen from Fig. 3 below, the frequency of home-based activity in ground truth data is much lower during early morning hours, when people are typically inactive. While, the frequency of work-based activity in ground truth is much higher during late night, which is probably because, people who work overtime tend to check-in at workplace more frequently, thus overestimating the frequency of work-based activity at night. Apart from these, the basic temporal distribution patterns of home-based and work-based activity from ground truth data are consistent with those from survey data. It means that, even as a small portion in the Weibo data, this ground truth data would be sufficient for representing the regularity of the temporal patterns of home-based and work-based activities and is scalable to the whole Weibo data as well.

It has to be mentioned that, the small size of ground truth data is largely due to the limitation of inferring home and work locations by POI category. Therefore, the labelling method introduced here is just for building the ground truth for training our proposed identification algorithm.

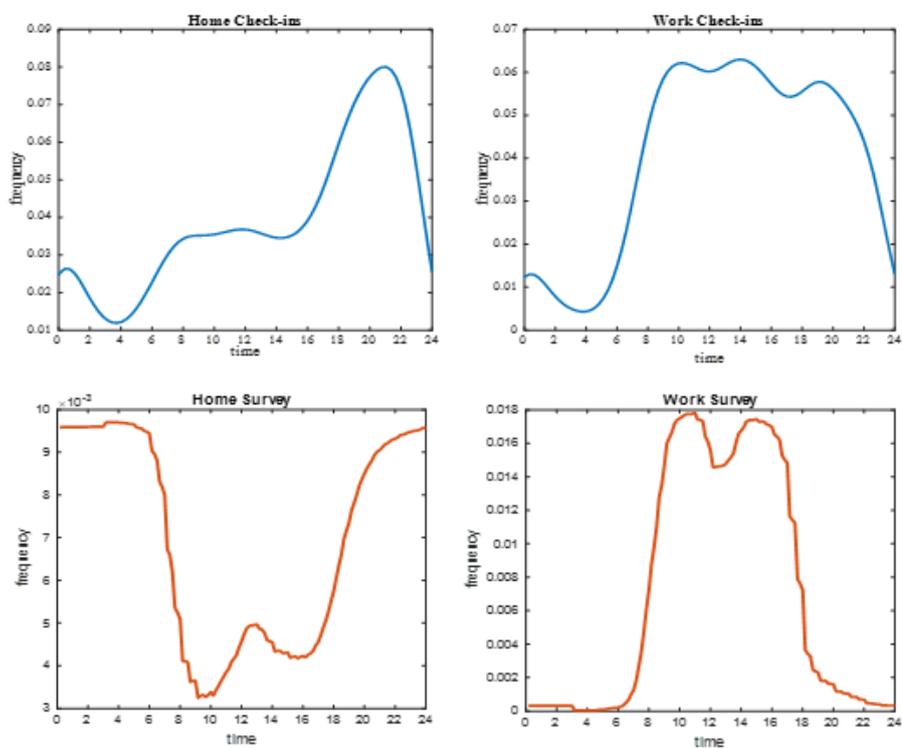

Fig. 3. Temporal pattern of home-based (left) and work-based activity (right) from ground truth (up) and survey data (bottom).

## 3.2 Purpose-specific location identification

### 3.2.1 Identification algorithm

We propose a threefold approach for home and work location identification adopted from Isaacman et al. (2011) on location identification with CDR, which has shown to be able to find users' important locations to within 3 miles 88% of the time. Existing studies have found strong agreement and robustness in human mobility patterns between CDR and check-in data (Cho et al., 2011; Lenormand et al., 2014). The factors used are listed below.

- ·   *Days*: The number of days on which the venue is geo-tagged.
- ·   *Cluster Days*: The number of days on which any venue in the cluster is geo-tagged. If two or more venues are geo-tagged on the same day, the day is counted only once.
- ·   *Timespan*: The number of days interval between the first and the last check-in with any venue in the cluster.
- ·   *Total Days*: The number of days between the first record and the last record in the dataset.
- ·   *Work Event*: The number of check-ins in the cluster on weekdays from 9am to 12pm and from 1pm to 6pm.
- ·   *Home Event*: The number of check-ins in the cluster on weekdays or weekends from 10pm to 7am.
- ·   *Work/ Home Event Percentage*: The percentage of '*Work/ Home Event*' of each cluster in all clusters.

We first spatially cluster the venues that appear in a user's check-in records. Because we do not have any prior knowledge regarding the number of clusters in each user's traces, which is probably different from one to another. As a result, we choose a distance-based clustering method (Isaacman et al., 2011; Chen et al., 2016). We first sort the unique venues in a user's check-in records in descending order based on '*Days*'. Relative importance of a certain venue in an individual's traces is represented by the days this venue is checked-in, instead of the total number of check-ins. Because, '*Days*' helps to decrease the influence of venues that are visited only on a few days, but that has a burst of events on those days (Isaacman et al., 2011).

Clustering starts with the first venue in the sorted list and makes this the centroid of the first cluster. For each subsequent venue, the algorithm would assign it to a corresponding cluster if it falls within a predetermined threshold radius, $r$, of any existing cluster centroid, or make this the centroid of a new cluster otherwise. When a new venue is added to any existing cluster, the centroid of that cluster will be moved to the days-weighted center of all the venues within the cluster. The algorithm continues until all the venues in the sorted list have been visited.

Then our identification algorithm runs to the second stage of filtering important clusters. Intuitively, important places in people's lives should be those that people visit frequently and recursively. A cluster will be assumed as important if it has been visited on more than a certain percentage of the total days, and the timespan in this cluster is more than a certain percentage of the total days. These translate to Eqs. (1)-(2) listed below. Here, $a$, $b$ are decision boundary values.

$$ClusterDays / TotalDays \geq a \tag{1}$$

$$Timespan / TotalDays \geq b \tag{2}$$

Finally, given all the important clusters, we infer where people live or work based on frequency of check-ins during pre-defined 'home' or 'work' event hours. As demonstrated in previous studies, for home identification, the most dominating factor is '*Home Event*' (Isaacman et al., 2011) or number of check-ins (Hossain et al., 2016). For workplace identification, there are two dominating factors; location with higher '*Work Event*' and lower '*Home Event*' is more likely to be one's workplace (Isaacman et al., 2011). Rules for home and work identification are illustrated as the flowchart in Fig. 4. Here, $k_1$ and $k_2$ are decision boundary values.

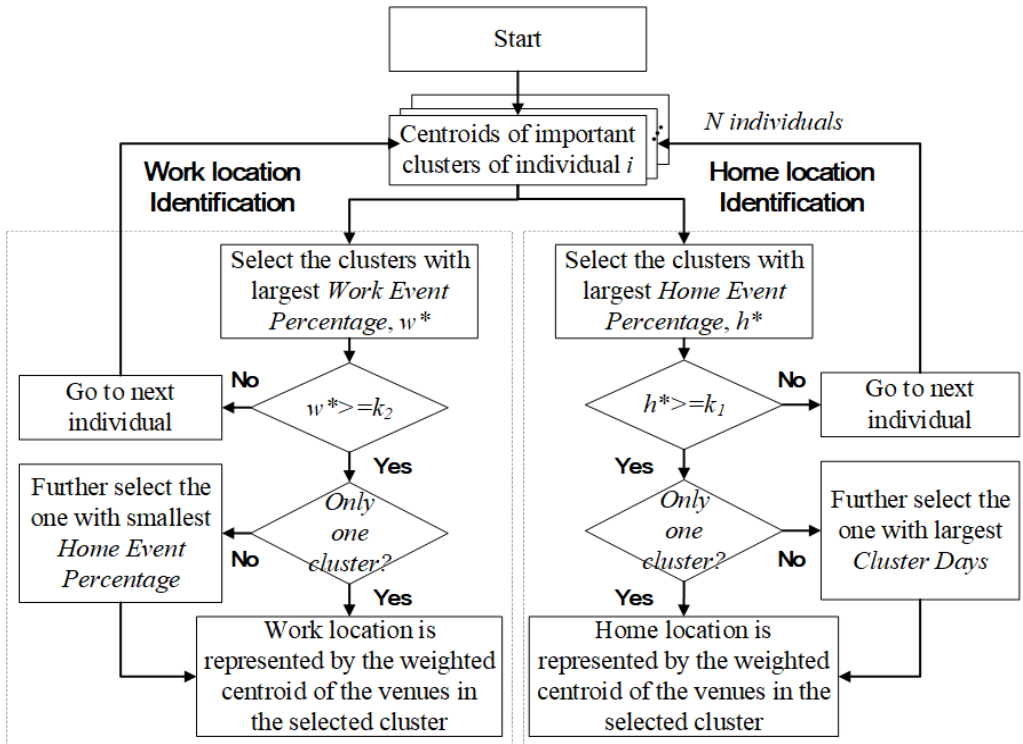

Fig. 4. Flowchart for identification of home location and work place.

The algorithm for entertainment and other place identification starts with check-in records of a user with home or workplace identified. For whom, our identification algorithm has already differentiated places where one visits on a regular temporal basis from those one visits flexibly. After removing geo-tagged locations in the clusters defined as '*Home*' or '*Work*', we can simply label each remaining location as '*Entertainment*', '*Other*' or '*Unknown*' based on their venue categories.

**3.2.2 Controlled random search algorithm**

We need to adjust the five tuning parameters, $r, a, b, k_1, k_2$, constantly until most individuals' resulting identified locations give the best possible fitting to the labelled ones in ground truth. A global optimization procedure called Controlled Random Search (CRS) is used for parameter calibration (Hendrix et al., 2002).

In the context of minimization problems, CRS algorithm generates $p$ uniformly distributed random points through $\Theta$, defined by the value ranges of parameter vector $\theta$, and stores the points and corresponding function values $(f_1,...,f_p)$. Let $f_{max}$ and $f_{min}$ be the largest and smallest function values in the storage, and $\theta_{max}$ and $\theta_{min}$ be the corresponding points. At each iteration, CRS randomly selects a subset $\{\theta_1,...,\theta_m,\theta_{m+1}\}$ of $m+1$ points from the storage. Then a new point $\theta^*$ is generated by reflecting a randomly selected point from the subset through the centroid of the other $m$ points. If the function value $f(\theta^*)$ at $\theta^*$ is smaller than $f_{max}$, then $\theta^*$ and $f(\theta^*)$ replace $\theta_{max}$ and $f_{max}$ in the storage respectively. Then the algorithm updates the corresponding values of $f_{min}$ and $\theta_{min}$ as well. The process stops when the maximum number of iteration stages $N$ has been reached.

The objective function is defined by Eqs. (3)-(5), which take both the identification accuracy and coverage into consideration, resulting in a number in [0, 1].

$$f = 1/2 \times \left( 1/{M^*}^2 \times \sum_{i=1}^{M^*} \frac{dis(H_i^*, H_i)}{d_{max}^H} + 1/{N^*}^2 \times \sum_{j=1}^{N^*} \frac{dis(W_j^*, W_j)}{d_{max}^W} \right) \quad (3)$$

$$d_{max}^H = \max(dis(H_i^*, H_i)), \quad i \in M^* \quad (4)$$

$$d_{max}^W = \max(dis(W_j^*, W_j)), \quad j \in N^* \quad (5)$$

Where, $H_i^*$ and $H_i$ represents the geo-coordinates of the labelled and identified home location of individual $i$ from ground truth. Similarly, $W_j^*$ and $W_j$ represents the labelled and identified work location of individual $j$ respectively. $M^*$ and $N^*$ are the number of users from ground truth whose home and work can be identified with our identification algorithm. $dis(H_i^*, H_i)$, $dis(W_j^*, W_j)$ calculate the spherical distance between identified and labelled locations, which are used as a measure of identification error. The error is normalized through dividing by its corresponding maximum value, $d_{max}^{H}$ and $d_{max}^{W}$. Mean distance error is then weighted by the reciprocal of *M\** or *N\** to compensate for the accuracy gains at the cost of low coverage.

In this study, clustering radius $r$ is set to be 20m, 50m, 70m, 100m, 200m, 300m, and 500m. Under each setting of $r$, we use CRS algorithm to find the best estimates of the other four parameters. By numerical test, the settings of $p$ and *N* in this study are 200 and 20,000 respectively.

**3.3 Sample reconstruction**

As is shown in Fig.2, there is a large variation of population representativeness between Weibo and survey data. Considering the strong influence of socio-demographics on people's activity participation and destination choices (Ma et al., 2014), the limitation of population coverage of Weibo data can only give us a biased understanding of human mobility. Therefore, we introduce a method for correcting population bias of check-in data by linking personal mobility patterns to these socio-demographics.

The basic idea behind sample reconstruction is to select an optimal configuration from the partial sample of Weibo data, according to the known distributions of characteristics, such as age, gender, from aggregated constraints calculated with travel surveys. Then the reconstructed sample is a more representative configuration of the real population, whose sociodemographic is aligning closely to that of survey respondents, while maintaining the rich variety of its original activity mobility patterns.

To proceed with a reliable sample reconstruction, an appropriate selection of constraint variables is crucial. First, the variables used need to be both included in the two data sources to enable the sampling process (Kirk, 2013; Ma et al., 2014). Second, they should cover or at least be strongly related to the variables of interest (Kirk, 2013). Kruskal–Wallis test, a non-parametric method for testing whether samples originate from distributions with the same median, is used to select attributes strongly related to travel behavior. Results show that, commuting distances vary significantly among people in different age groups and among people living in different regions (*c.f. SI* Table 2). People aged 30~40, and people living in suburban areas tend to commute further. While, gender and education background only have slight influences on individuals' commuting distances. In this study, we conduct sample reconstruction for commuters and non-commuters separately to account for the influence of employment status on individuals' mobility patterns. For each case, we use the tabulations of age, gender, and home district, as well as the cross-tabulations of age and gender at the TAZs level as constraints. Home district is defined as the district that Weibo users' identified home locations belong to. Besides, the difference between commuters and non-commuters are determined by whether their work locations can be identified or not.

Flexible Modelling Framework (FMF), developed and tested at the University of Leeds since 2005, is an open source software framework for spatial population synthesis (Ma et al., 2014). It incorporates a static spatial simulation algorithm based on Simulated Annealing, which proves to be one of the most popular and effective methods for generating synthetic spatial microdata at different geographical scales (Kirk, 2013). In addition to the FMF population synthesizer, the traditional IPF and Combinatorial Optimization (CO) methods are still very popular (though they have drawbacks), especially in those cases where a simple and easy population synthesizer is adequate. The choice of FMF in our study is largely due to the availability of the open source software package that is relatively feasible to implement.

Simulated Annealing approach is executed for each TAZ individually. Individuals from each zone will first be given a unique person ID. The algorithm then clones, adjusts, and tests individuals in the generated population repeatedly until an appropriate set of individuals is established whose tabulated constraint variables give the best possible fit to the target aggregated sums. The goodness-of-fit level is measured as *Total Absolute Error* (Eq. (6)).

$$TAE = \sum_{i=1}^{N} \sum_{j=1}^{M} |T_{ij} - E_{ij}| \qquad (6)$$

Where, $T_{ij}$ is the count of individuals in category $j$ living in zone $i$ in survey data; $E_{ij}$ is the count of individuals in category $j$ living in zone $i$ in the generated population. *N* is the number of TAZs. $j$ refers to one specific category in Table 1. *M* is the total number of categories.

Table 1 Constraint configuration for commuters/non-commuters

| Constraint Variables | Categories |
| --- | --- |
| Gender | Male (m), Female (f) |
| Age | 15-24, 25-26, 27-28, 29-30, 31-33, 34-39, 40-49, 50+ |
| Age by Gender | 15-24-f, 25-26-f, 27-28-f, 29-30-f, 31-33-f, 34-39-f, 40-49-f, 50+-f<br>15-24-m, 25-26-m, 27-28-m, 29-30-m, 31-33-m, 34-39-m, 40-49-m, 50+-m |
| Home District | Dongcheng, Xicheng, Chaoyang, Haidian, Fengtai, Shijingshan, Changping, Shunyi, Tongzhou, Daxing |

After sample reconstruction, activity-specific location choices of Weibo samples can be attached to the generated population based on the unique person ID.

## 3.4 Out-of-sample validation

Here we select three indexes for evaluating the performance of our method, through out-of-sample validation with survey data, regarding spatial distribution pattern of activity-specific locations and distribution of commuting distance. In travel surveys, respondents' activity locations are mostly recorded using the TAZs code since GPS devices are rarely used to track the precise locations of individuals during surveys. Therefore, in this study, we use TAZ as the geographical unit for cross-validation between two data sources.

*Cosine Similarity* (*CS*) is a measure of similarity between two non-zero high-dimensional vectors of an inner product space that measures the cosine of the angle between them (Wikipedia, 2017). In our study, each TAZ represents a unique dimension, and the region is characterized by a vector where the value of each dimension corresponds to the percentage of users with identified purpose-specific places located in corresponding TAZ. *CS* then can be used to measure how similar two spatial distribution maps are likely to be, which is defined as Eq. (7). *CS* takes the value in [0, 1]. A larger *CS* indicates a better fit.

$$CS = \frac{\sum_{i=1}^{N} p_i^C \times p_i^S}{\sqrt{\sum_{i=1}^{N} \left(p_i^C\right)^2} \times \sqrt{\sum_{i=1}^{N} \left(p_i^S\right)^2}} \quad (7)$$

Where:

$p_i^C$: The percentage of Weibo users with identified purpose-specific places located in TAZ *i*;

$p_i^S$: The percentage of survey respondents with recorded purpose-specific places located in TAZ *i*;

$N$: The number of TAZs.

As a supplement to *CS*, we also calculate the *Distance between Centers of Gravity* (*DC*) of identified locations with check-ins and recorded locations in surveys. In this case, the *Centers of Gravity* of purpose-specific activity distribution maps is defined as Eq. (8). The spherical distance between two centers is calculated from their longitudes and latitudes.

$$Centers = \left( \sum_{j=1}^{n} X_j \times \frac{N_j}{\sum_{i=1}^{n} N_i}, \ \sum_{j=1}^{n} Y_j \times \frac{N_j}{\sum_{i=1}^{n} N_i} \right) \tag{8}$$

Where:

$X_i$: The latitude of centroid of TAZ $i$;

$Y_i$: The longitude of centroid of TAZ $i$;

$N_i$: The number of activity-specific destination choices located in TAZ $i$ summed across all the individuals, and $n$ is the number of TAZs.

We further compare the distribution of commuters by commuting distance traveled (one-way) from residence location to workplace, for Weibo's identified commuters and survey's self-reported commuters. Commuting distance is calculated as the spherical distance between individual's residence and work location. *Coincidence Ratio* (*CR*) measures the percent of the area that 'coincides' for the two curves of distribution to compare (Yang et al., 2015), as defined in Eq. (9). The maximum commuting distance in our dataset is about 120 km long, thus the length interval is identified as 0.24 km, which results in 500 intervals. *CR* ranges from 0 to 1, with a higher value indicating a better fit.

$$CR = \frac{\sum_{i=1}^{N} \min\left(p_i^C, p_i^S\right)}{\sum_{i=1}^{N} \max\left(p_i^C, p_i^S\right)} \quad (9)$$

Where:

$p_i^C$: The percentage of commuting distance in interval $i$ in check-in data;

$p_i^S$: The percentage of commuting distance in interval $i$ in travel log survey;

$N$: The number of intervals.

**4 Results**

**4.1 Activity-mobility pattern**

The calibration results show that, all of the four parameters, concretely, $a, b, k_1, k_2$ give the best converging performance at the radius of 300 m (*c.f.* *SI* Figure 1 and *SI* Table 3), which is thus set to be the radius for our clustering algorithm; and, the best estimates of $a, b, k_1, k_2$ are set to be the means of the best 200 points in storage, which are 0.93%, 9.25%, 6.1%, and 1.92% respectively.

Running the calibrated identification algorithm on the whole Weibo dataset, generates a newly labelled sample in which 65,450 users' home and 63,399 users' work place can be identified. The *CS* index for home spatial layout is 0.48. Hot spots for the residences of Weibo users are mostly distributed in the districts of Haidian and Chaoyang, where most schools and universities are located; whereas, hot spots for survey respondents' home locations are distributed in inner area and southern suburban regions (*c.f.* Fig. 5 $a_1$ and $a_2$). This is probably because Weibo users are mainly students, whose major residences are within the 'School District'. A high similarity of 0.65 for spatial distribution exists between Weibo users' and survey respondents' workplaces (*c.f.* Fig. 5 $b_1$ and $b_2$). In Beijing, the city center is still attractive regarding work opportunities and commercial infrastructure, where most employment is concentrated now, which may account partially for the strong coincidence between work distributions discerned from the two datasets. Compared to commuting locations, less consistency appears in the spatial distributions of locations for entertainment and other activities, which is around 0.30~0.35. As inferred from Fig. 5 $c_1$ and $c_2$, Weibo users' leisure areas are mostly towards the districts of Dongcheng and Chaoyang, whereas survey respondents tend to go to the inner areas for entertainment. This is probably because Weibo users, representative of young population in a city, prefer commercial facilities such as shopping centers and bars that mostly located in the north and east of Beijing to public parks in the city center. Intuitively, one would not travel long distances to a shopping mall far away from home, when similar facilities are available nearby. Therefore, it is highly possible that most places where people participate in non-commuting activities are distributed around their living places. Therefore, hot spots for non-commuting activities are concentrated in the same region where most people live (for other density maps, see *SI* Figure 2).

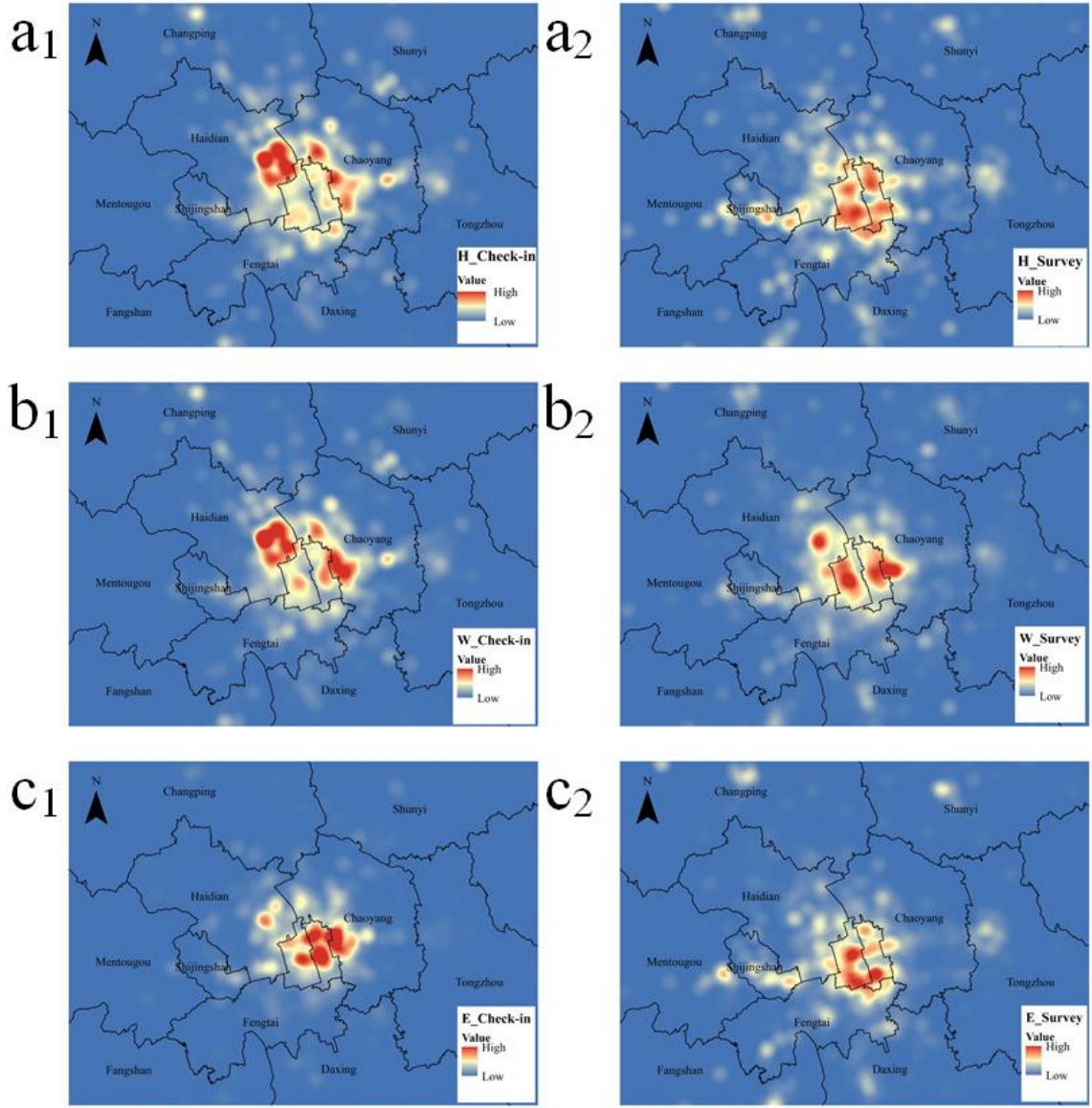

Fig. 5 Validation of activity-specific locations identified from check-ins with independent travel logistic surveys. The base map shows the spatial layout of Beijing's administrative districts. Kernel density estimation is conducted to obtain a smooth distribution. ($a_1$ and $a_2$) The identified-home and recorded-home density maps for major metropolitan districts in Beijing. ($b_1$ and $b_2$) The identified and recorded workplace density maps of Beijing. ($c_1$ and $c_2$) The identified and recorded entertainment density maps of Beijing.

We further evaluate the commuting distance distribution of identified local commuters. Obvious difference exists between the two cumulative distribution curves, with the *CR* index of only 0.28; check-in data overestimate the frequency of zero-distance commuting (*c.f.* Fig. 7). In reality, some people may work at home or live at their workplace. For those individuals, our algorithm would end up with the same location identified as both home and workplace. The facts that students constitute a significant portion of Weibo users, and home-based workers would not be treated as commuters in travel surveys, are probably the reasons for this mismatch.

**4.2 Sample reconstruction**

We perform sample reconstruction on the central three zones of inner area, functional extended area and new urban district to represent urban Beijing, because, as shown in our dataset, these zones account for 91% of all the population and over 99% of working opportunities of Beijing. Fangshan district is excluded for the limited number of sample data there. Based on the constraint variables we specify, only the Weibo samples whose age and gender information are available, and home locations can be identified as well as located within the selected districts, would be used for sample reconstruction. These specifications give us a dataset composed of 24,518 samples.

Our analysis shows that all the constraining features at the TAZ level are reproduced with no misclassification (*c.f. SI* Table 4). The reconstructed population can be seen as an artificial sample as representative as survey respondents under the constraints we formulate. The results show that socio-demographics and commuting patterns of the reconstructed population matches well with those of survey respondents, with the *CS* index for the spatial distribution of population and workplace as 100% and 65% respectively. A good match is achieved for the spatial distributions of the destinations for entertainment (54%) and for other activity (54%). Besides, all the centers of gravity migrate southwest after sample reconstruction (*c.f.* Fig. 6 b); the resulting new activity-specific centers move closer towards the corresponding centers discerned from survey data (*c.f.* Fig. 6 a).

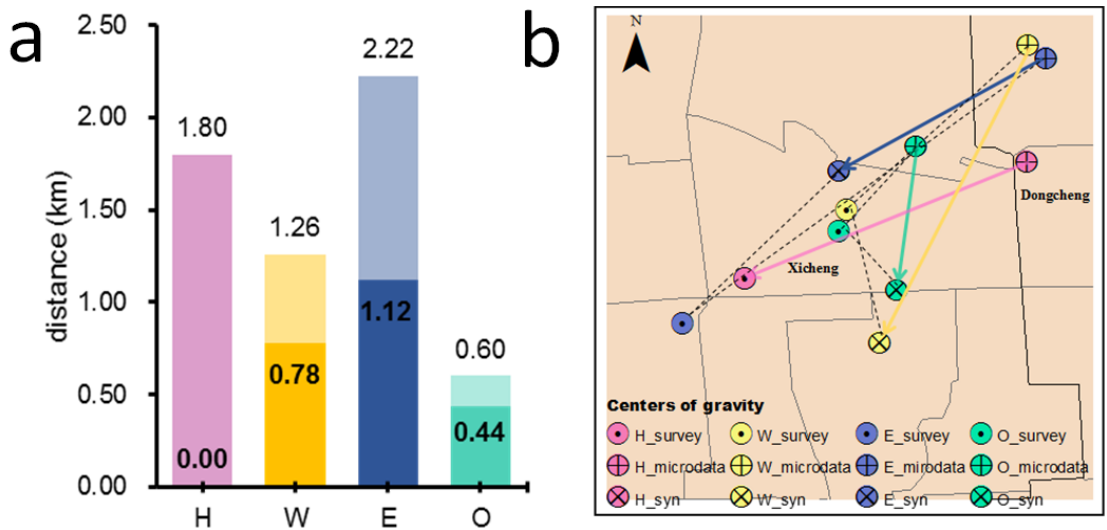

Fig. 6 Performance of sample reconstruction. (a) Distance between Centers of Gravity index before and after sample reconstruction for activity category of Home (H), Work (W), Entertainment (E) and Other (O). The transparent bars with un-bold data labels indicate the check-in data's baseline accuracy, expressed in terms of Weibo microdata-survey similarity. (b) Centers of Gravity of activity-specific spatial distribution maps. We refer to the check-in data as 'microdata', and reconstructed data as 'syn'. Movements of centers are denoted by arrows. Here, synthetic center coincides with survey center for home distributions.

Generally, our proposed sample reconstruction technique for correcting sample bias works relatively better for commuting activities than for non-commuting activities. There are two main reasons. First, working opportunities in Beijing are center-towards; meanwhile, the age profiles of commuters are normally concentrated within a specific age range, like 27 to 50, which is covered relatively sufficiently in both datasets. For non-commuting activities, the active participants are usually people aged 50 and above, who have more leisure time available after retirement. However, these people are under-represented in check-in data. Reconstructing a population with 'unbiased' age profiles will result in cloning just a limited number of elderly persons multiple times, which could only bring small modifications to the resulting activity spatial distributions. Second, compared with commuting activities, individuals usually have much more freedom when choosing recreational destinations. Therefore, it is quite likely that the discrepancies result from individual diversities.

Furthermore, we examine at the sub-district level how the commuting distance distribution patterns change after sample reconstruction. In general, as compared with identified commuters from Weibo data, the commuting distance distribution curve of synthetic commuters from reconstructed population is more consistent with that of survey respondents (*c.f.* Fig. 7). More specifically, after sample reconstruction, the *CR* index is increased substantially from 0.23 to 0.63. Furthermore, in inner zone and functional extended zone, the district-specific *CR* index of the synthetic commuters are all approximately 0.50 or above (*c.f. SI* Table 5). Commuting distance distributions of commuters from new urban districts are only improved to a limited degree. This is probably because in these areas, much bigger diversity exists in residents' travel behaviors, while available data there may not be sufficient to reflect fully this large diversity.

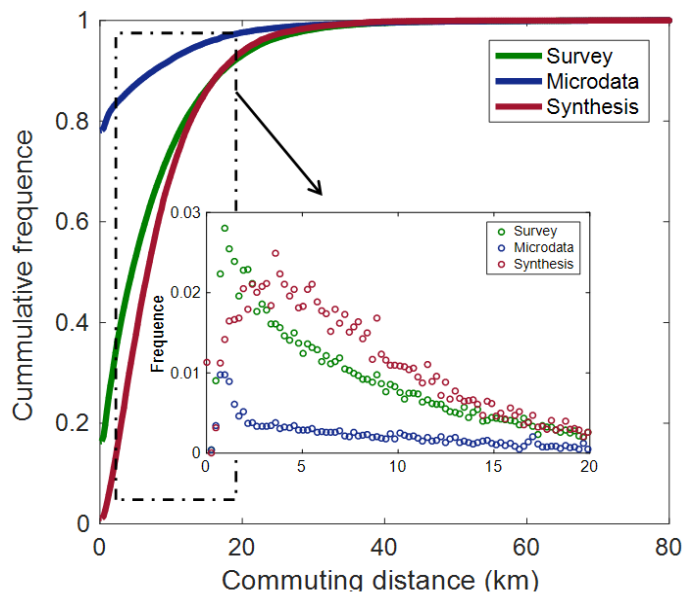

Fig. 7 Commuting distance distributions. The direct and cumulative distribution curves of commuting distance. Here, 'Survey' refers to survey data; 'Microdata' refers to Weibo data after location identification; 'Synthesis' refers to generated population after sample reconstruction. After sample reconstruction, the frequency of zero-distance commuting has been largely reduced. The synthetic distribution curve is more consistent with survey data though it slightly overestimates the frequency of commuting distance longer than 3km.

Based on the above analysis, we have demonstrated that sample reconstruction can be a useful sampling bias correction method to adjust the bias in social media data. However, discrepancies still exist between spatial mobility patterns discerned from synthetic and survey data. First, there are still other socio-demographics assumed to be influential on travel behavior, such as employment, occupation, housing area, and housing tenure (Ma et al., 2014), which are absent in Weibo users' profiles and thus cannot be included in our synthesis process. As a result, the generated population may not match well with survey respondents under these attributes. Second, check-in data and travel logistic survey data are both a representative sample from the total population and each just reflects one component of the entire array of complicated daily human behavior. Considering that the data we use are only a subset of all geo-tagged check-in records available on Weibo, when a sufficiently large dataset is utilized, the inferred results would improve further.

# 5 Conclusion and Discussion

In this research, we explored the capacity of social media data for human mobility analysis. A three-step methodological framework, including purpose-specific location identification, bias modification, and out-of-sample validation, was proposed and then tested using check-in records crawled from Sina Weibo. An independent citywide travel survey was used as the benchmark dataset for validating the results. Our analysis shows that Weibo users' and survey respondents' demographic profiles and mobility patterns have distinct features. This implies a strong influence of socio-demographics on people's activity participation and destination choices. We first tested the possibility of using a spatial population synthesis technique to solve the bias-modifying problem of check-in data. Our results showed that socio-demographics and activity-specific spatial layouts of the reconstructed population match satisfactorily with those of survey respondents. Comparing the inferred results from big data to aggregated results from an independent travel survey represents a very important step towards big data validation. However, it should be noted that the inferred results at the individual level could have great uncertainty even though a high level of accuracy is observed at the aggregate level (Chen et al., 2016). Besides, the target distribution used for sample reconstruction in our study was extracted from the survey data. The extent to which the resulting synthetic population could represent the real one largely depends on the quality of the survey data. In general, such survey data have a relatively small sample size, compared with the whole population, and thus the data might not be representative enough in some cases. Another thing to keep in mind is that, check-in data and travel log survey data are both sampled from the total population, and each only presents a certain configuration of the reality. Therefore, combining knowledge learned from multi-source data would lead us much closer to the real system. The method we proposed in this research would also be a straightforward and easy-undertaking technique for cross-domain integration of heterogeneous human spatiotemporal data.

Another remaining challenge of using check-in data for transport-related analysis, especially in advanced travel demand models, is that it has missing activities, since users share their activities voluntarily. Moreover, information on the mode of transport is also missing. However, these are not the focus of this article. An attempt in this domain can be found in Hasan and Ukkusuri (2018), where they applied Continuous-Time Bayesian Network model to extract the true transition and duration distributions from the incomplete trajectories of Twitter users. Transportation mode can also be extracted from Weibo text using text mining and natural language processing approaches. As suggested by Rashidi et al. (2017), constructing a dictionary for this purpose is not as complicated.

Social media check-in data opens up a novel dataset for urban analysis. Compared with traditional surveys, this dataset apparently has more advantages, including larger sample size, real-time updating, less-prone to cheating, and a much lower acquisition cost. Especially in new urban districts, where population size, structure, and mobility patterns are significantly changing, whereas the survey data is not available or too expensive to collect, this study has validated that through a proper way for analyzing Weibo data, such data can be a moderate and cost-effective alternative source of mobility information. The proposed framework can be extended in several directions. Relying on the commuting mode shares and emission intensities of each mode from supplementary databases, the commute estimates in our work can be easily extended to carbon footprints analysis. Another potential extension is to fill the missing activities in check-in sequences with the help of activity transition probabilities and purpose-specific trip distance distributions from travel surveys. Such models will provide valuable information for transportation emission modelling.

For instance, the activity-location sequences generated can be integrated with state-of-the-art network assignment model to obtain dynamic traffic flows on the road network, and then feed them to transport emission model, which will be able to infer the gas consumption and pollution emissions incurred by passenger travel at a high temporal and spatial scale. Most importantly, Weibo data makes a detailed activity-based travel analysis possible as geo-tagged venues are categorized based on the POI type of the visited places. Incorporating such information enables urban planners and environmental specialists to estimate the volume of air pollutants that would be generated during certain time intervals, between specific regions, and to fulfill different needs. This can facilitate a better understanding of urban organization and its environmental effects. The longitudinal nature of social media data could also enable planners to deal with valuable questions like 'how does human mobility evolve over a year?', 'what is the role of the underlying built environment in forming human mobility and further transportation emission?'

**Acknowledgements**

We would like to thank Fuzheng Zhang, Jianxun Lian, and Danyang Liu from Microsoft Research Asia for helping us crawl and manage Weibo check-in data.

# Appendixes

Appendix A

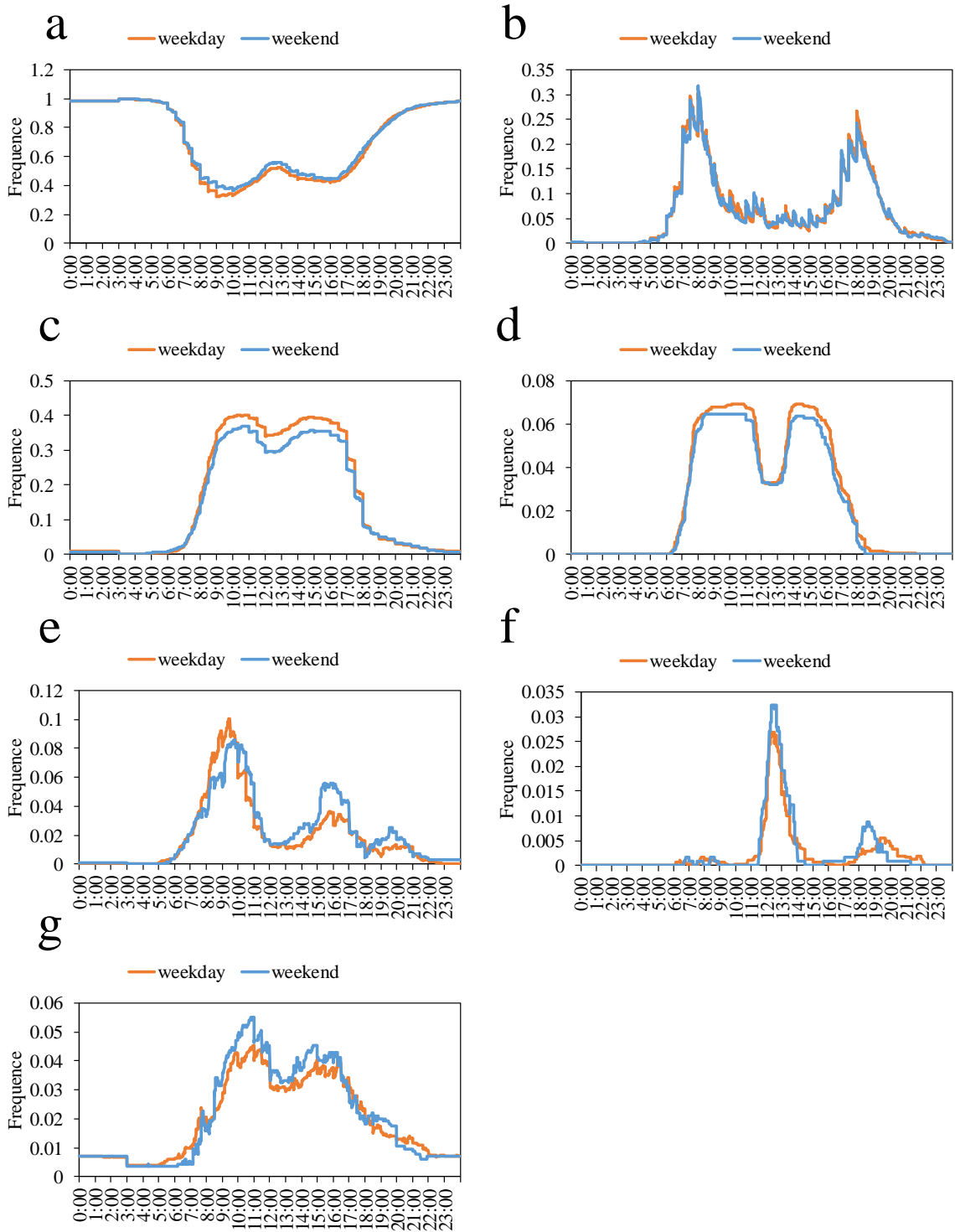

Appendix Figure 1. Temporal features of main activity category. (a) Home; (b) Travel; (c) Work; (d) School; (e) Entertainment; (f) Eating out; (g) Other.

# Supporting Information for
## Social media and mobility landscape: uncovering spatial patterns of urban human mobility with multi source data

In this supporting information, we present the essential extended details of the results and analysis.

## Calibration Results

This section gives the results of calibrating home and work identification algorithm with the ground truth data, given the optimal objective. Besides, we also evaluate the calibrated results based on mean distance error, standard deviation of distance error and identification coverage (that is, the share of users with home/work identified in those with home/work labelled from the ground truth) (*c.f. SI* Table 3).

After 20,000 times of iteration, the objective functions under seven settings of clustering radius all converge to some single value. It means that the pre-set algorithmic control parameters of CRS (including, iteration stages $N$ and initial points in storage $p$) are sufficiently moderate to generate a set of parameters with which most individuals' resulting identified home and workplaces give the best fittings to the labelled ones.

We further calculate the 25th, 50th, 75th, 97.5th percentile error between the home and workplaces as identified by our algorithm and the labelled ones as in ground truth. Both algorithms perform well, achieving median errors of 0.08 km and 0.64 km, respectively. Moving out to the 75th percentile, the home algorithm continues to work well, with 2.90 km of error, whereas the work algorithm's error increases to 6.58 km. A further exploitation into the few cases with larger distance errors reveals that the poor performance usually happens to users who do not use Weibo frequently at workplace, or users who would post a microblog only when they are working overtime, which time period is not covered by the peak work event hours defined in our algorithm. In general, our home algorithm performs better than work algorithm.

## Sample Reconstruction

It has to be noted that, in this study, due to the limited size of non-commuters identified with check-in dataset, population reconstruction for those has to sample from the whole microdata. Considering the dataset we use is only a subset of all geo-tagged check-in records available on Weibo, when a sufficiently large dataset is utilized, this process would improve further.

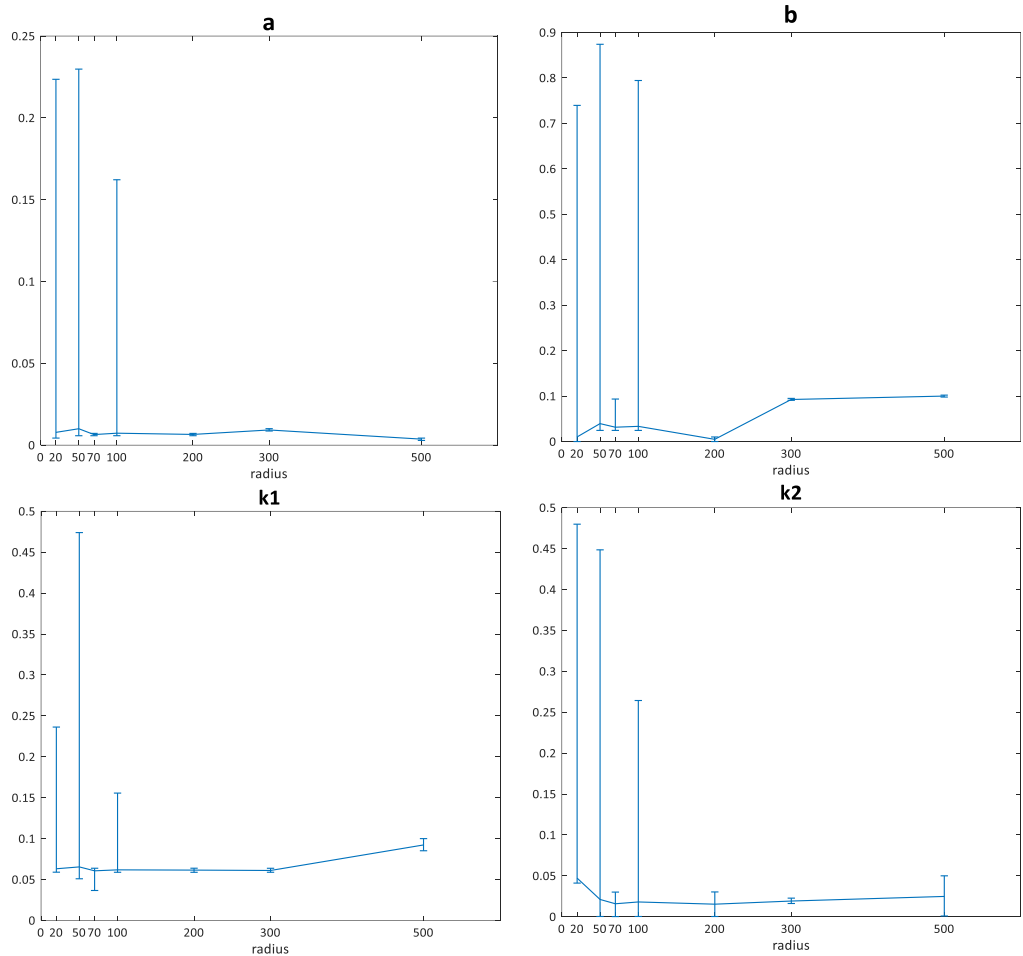

**Figure 1 The first 200 best parameter estimates of CRS algorithm under different settings of clustering radius (m).** The upper bond: indicates the maximum value, lower bond, the minimum value and center, the best estimate of objective function. Comparatively, all of the four parameters give the best converging performance at the radius of 300m.

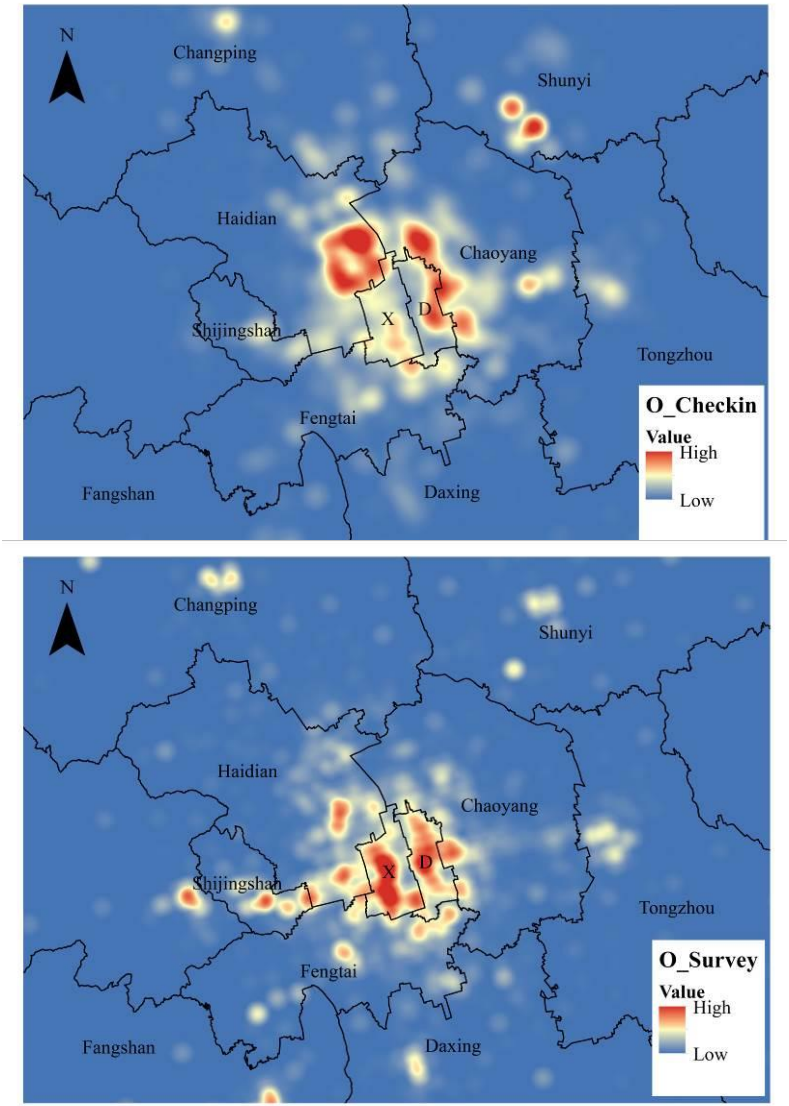

**Figure 2 Other activity density maps.** The identified (up) and recorded (down) other activity density maps for the major metropolitan area in Beijing (person_ days/km$^2$), with a *Cosine Similarity* index of 0.35. Among all the non-commuting locations, we calculate the number of days each unique venue is visited (*Days*). Then, the popularity of a certain place as a candidate location for other activity can be calculated as the sum of the corresponding *Days* of all the users.

Table 1 Bags of words used when building ground truth data

|  | Words |
|---|---|
| Work/School: | 上班/下班/上学/放学(work/school), 单位/公司(company); 办公室(office room); 工位(working station); 教室(classroom); 加班(work overtime); 值班(on duty) |
| Home: | 家(home), 床(bed), 起床/早起(wake up), 睡觉(sleep), 早安(morning), 晚安(good night), 宿舍/寝室(dormitory); 出门(go out) |

Table 2 Mean values of commuting distances by attributes

|  | Attributes | Distance(km) |  | Attributes | Distance(km) |
|---|---|---|---|---|---|
| Gender | female | 2.20 | Education | primary | 2.01 |
|  | male | 2.02 |  | secondary | 2.10 |
|  |  |  |  | tertiary | 2.13 |
|  | Chi-sq (P) | 0.01(0.91) |  | Chi-sq (P) | 0.00(0.99) |
| Age | 15~24 | 1.81 | District | Dongcheng | 1.89 |
|  | 25~26 | 1.67 |  | Xicheng | 1.53 |
|  | 27~28 | 2.12 |  | Chaoyang | 2.07 |
|  | 29~30 | 2.53 |  | Haidian | 1.39 |
|  | 31~33 | 2.66 |  | Shijingshan | 3.27 |
|  | 34~39 | 2.58 |  | Fengtai | 2.82 |
|  | 40~49 | 1.86 |  | Changping | 3.03 |
|  | 50+ | 0.93 |  | Shunyi | 3.25 |
|  |  |  |  | Tongzhou | 3.49 |
|  |  |  |  | Daxing | 2.50 |
|  | Chi-sq (P) | 335.92(0.00) |  | Chi-sq(P) | 103.64(0.00) |

Info: Kruskal–Wallis test (or, KW test), a non-parametric method for testing whether samples originate from distributions with the same median, is used to select attributes strongly related to travel behavior. A significant KW test, with a P value close to zero, indicates that at least one sample is from a distribution with a different median.

Table 3 Calibration results of home and workplace identification

| r(m) | Home | | | Work | | | Objective Function |
|---|---|---|---|---|---|---|---|
| | Mean | Standard Deviation | Coverage | Mean | Standard Deviation | Coverage | |
| 20 | 3.32 | 6.83 | 91% | 4.60 | 7.97 | 87% | 4.28E-04 |
| 50 | 3.24 | 6.76 | 88% | 4.42 | 7.87 | 86% | 4.37E-04 |
| 70 | 3.20 | 6.63 | 95% | 4.69 | 7.76 | 93% | 4.38E-04 |
| 100 | 3.18 | 6.64 | 92% | 4.66 | 7.90 | 92% | 4.34E-04 |
| 200 | 3.43 | 6.89 | 97% | 4.66 | 7.29 | 96% | 5.05E-04 |
| 300 | 3.18 | 6.54 | 92% | 4.48 | 7.25 | 91% | 5.08E-04 |
| 500 | 3.67 | 7.32 | 97% | 5.21 | 7.87 | 99% | 5.03E-04 |

Table 4 Representation of the model constraint at the TAZs level

| | Commuting | | Non-commuting | |
|---|---|---|---|---|
| | TAE | CPE (%) | TAE | CPE (%) |
| District | 0 | 0 | 0 | 0 |
| Age | 0 | 0 | 0 | 0 |
| Gender | 0 | 0 | 0 | 0 |
| Age by Gender | 0 | 0 | 0 | 0 |

Info: Total Absolute Error (TAE) and Cell Percentage Error (CPE) are used to evaluate the goodness-of-fit of the synthetic population. TAE is calculated with Eq. (6) in the manuscript, while CPE is derived by TAE/N*100, where N is the population of the relevant cell, zone or attribute.

Table 5 Sub-district level *Coincidence Ratio* of commuting trip-length distribution before and after sample reconstruction

| | Sample | Synthesis | Sample size |
|---|---|---|---|
| **Total** | **0.23** | **0.63** | **24,518** |
| Dongcheng | 0.24 | 0.59 | 1,589 |
| Xicheng | 0.20 | 0.58 | 2,137 |
| Chaoyang | 0.25 | 0.55 | 6,979 |
| Haidian | 0.23 | 0.52 | 5,820 |
| Fengtai | 0.24 | 0.60 | 2,671 |
| Shijingshan | 0.20 | 0.49 | 714 |
| Changping | 0.28 | 0.43 | 1,698 |
| Shunyi | 0.26 | 0.25 | 977 |
| Tongzhou | 0.22 | 0.41 | 925 |
| Daxing | 0.20 | 0.42 | 1,008 |